\documentstyle[aps,multicol,epsfig]{revtex}    

\newcommand{\beq}{\begin{equation}}   
\newcommand{\eeq}{\end{equation}}   
\newcommand{\bea}{\begin{eqnarray}}   
\newcommand{\eea}{\end{eqnarray}}      

\begin{document}      

\title{Noise Correlations in Shear Flows} 

\author{ Bruno Eckhardt$^1$ and Rahul Pandit$^2$\cite{byjnc}  }      

\address{$^1$Fachbereich Physik, Philipps-Universit\"at Marburg, 
D-35032 Marburg, Germany}     

\address{$^2$Centre for Condensed Matter Theory, Department of Physics,  
Indian Institute of Science, Bangalore - 560 012, India} \date{\today}  

\maketitle{ }       

\begin{abstract} 
We consider the effects of a shear on velocity fluctuations in a flow. 
The  shear gives rise to a transient amplification that not only influences 
the  amplitude of perturbations but also their time correlations. We show 
that, in  the presence of white noise, time correlations of transversal 
velocity  components are exponential and that correlations of the  
longitudinal components are exponential with an algebraic prefactor. 
Cross correlations between transversal and downstream components  
are strongly asymmetric and provide a clear  indication of non-normal
amplification. We suggest experimental tests of our  predictions.\\  

47.27.-i, Turbulent flows, convection, and heat transfer \\ 
05.40.Ca, Noise  \\ 
47.20.Ft  Instability of shear flows   \\ 

\end{abstract}         


\begin{multicols}{2}

\section{Introduction}  
There has been a resurgence of theoretical interest in shear-driven flows, 
 ubiquitous both in nature and in the laboratory \cite{Grossmann}, because 
 the  linearization of the Navier-Stokes equation about a laminar shear 
 profile  gives rise to a non-normal linear operator. The eigenstates of this 
 operator  are not orthogonal so, even if the laminar profile is asymptotically 
 stable,  perturbations can grow as transients for a while before submitting 
 eventually  to viscous damping
 \cite{Grossmann,Landahl,Boberg,Trefethen,Farrell1}. 
This  effect occurs generically if non-normal linear operators arise during a  
linear-stability analysis and it has been discussed at length in various  
contexts 
\cite{Ecology}, including the hydrodynamical one we concentrate on here.   

Several authors have advocated the use of peudo-spectra in problems
involving non-normal operators \cite{Trefethen,SH}. A complementary 
approach is to  characterise the behaviour of non-normal systems by studying 
their response to  an externally imposed noise. Farrell and Ioannou 
\cite{Farrell2,Farrell3}  showed that, in general, this results in an increased 
variance; they determined  amplification rates and the perturbations that give 
rise to the largest  amplitudes for shear flows \cite{Farrell4}. Rather little 
has been said about  temporal correlations in such systems; an important, early 
exception is the study by 
Onuki\cite{Onuki} whose emphasis is quite different from ours as we  explain 
below.  But the temporal relation between lift-up and other 
instabilities is at the heart of the steady cycle that Waleffe et al
propose as the main mechanism for sustaining turbulent fluctuations in 
shear flows \cite{Hamilton,Waleffe1,Waleffe2}. The idea is that downstream 
vortices give rise to downstream streaks by non-normal amplification, 
that the streaks go unstable to vertical vortex formation and that
finally these vertical vortices are folded over into downstream vortices by the 
linear shear profile. 

It is our aim here to present results on correlations in  linear, non-normal 
systems driven by white noise and to discuss their  relevance to perturbed 
shear flows. Specifically, we will show that temporal cross  correlations
 between a {\it streak} and a {\it vortex} component can provide  
 unambiguous signs of lift-up and non-normal amplification.

The formalism we use to study the fluctuations is similar to 
{\it Rapid Distortion Theory} or RDT \cite{RDT,Nazarenko}. 
We assume that we have a prescribed strong shear and follow a 
perturbation in Fourier space, using Kelvin modes. In order to 
avoid technical difficulties related to advecting frames of reference 
or time-dependent wave vectors, we use perturbations without variation 
in the flow direction. Nevertheless, as we will discuss later on, there is 
numerical evidence that this does not significantly affect the main 
conclusion that we draw about cross-correlations.

 The outline of the paper is as follows: In Section \ref{vortex_streak} we 
  discuss a stochastic model with two degrees of freedom, with one vortex 
  and one shear component, that highlights the non-normal coupling and 
  the temporal  correlations to be expected. We then turn in Section 
  \ref{Fourier} to a full  discussion of temporal correlations in 
  the stochastically forced Navier-Stokes  equation linearised about a 
  laminar shear profile. Section \ref{conclusions}  contains 
  concluding remarks, suggestions for experimental tests of our 
   predictions, and a discussion of the relation of our work 
   to earlier studies.  

\section{A Simple Model} 
\label{vortex_streak} 
\subsection{Vortex-Streak Coupling in Shear Flows} 
Consider a linear shear profile, ${\bf U}_0= S z {\bf e}_x$. Coordinates
are chosen in the meteorological convention, with $x$ pointing downstream,
$y$ in the spanwise direction and $z$ pointing in the direction 
of the shear. The Navier-Stokes equation for the fluid velocity 
${\bf u}\/$ linearised  around this flow is 
\beq 
\partial_t {\bf u} + ({\bf u}\cdot\nabla) 
{\bf U}_0 + ({\bf U}_0\cdot\nabla) {\bf u} 
= - \nabla p + \nu \Delta {\bf u}\, , 
\label{NS_lin} 
\eeq 
where $p\/$ is the kinematic pressure and $\nu\/$ the kinematic viscosity of  
the fluid. To keep the analysis as simple as possible we work with the Fourier  
modes appropriate for periodic boundary conditions in spanwise and 
downstream  directions, and free-slip boundary conditions on two 
parallel planes in the  
normal direction. The analysis of the linear problem with Kelvin modes  
\cite{Kelvin,Craik,Marzinzik} shows that modulations in the downstream  
direction give rise to a time-dependent wave vector and 
faster-than-exponential damping. Farrell and Ioannou \cite{Farrell4} 
also show that the most important  
modes for non-normal amplification do not have a downstream variation.  
Therefore, we consider only perturbations with wave numbers  
${\bf k} = (0, k_y, k_z)$, where $k_y\/$ is continuous and 
$k_z=\pi n/d$, with  $n\/$ an integer and $d\/$ the distance 
between the bounding planes. These  considerations help us to 
identify scales for the flows considered in this section; 
their full significance will become clear in the next section.  

\begin{figure}
\centerline{\epsfig{file=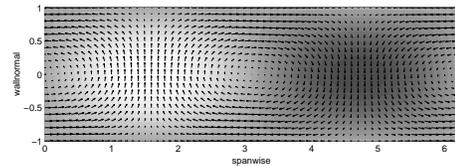,angle=0,width=6cm}} 
\caption[]{ Sketch of a vortex and a streak in a shear flow, 
looking downstream. 
The velocity components of the vortex lie in the plane and 
are indicated by the arrows. 
The  streak has only one velocity component perpendicular to 
the plane; its values are 
indicated by a grey scale. The extrema of the streak (light and dark regions) 
are located near the maximal normal velocities. All scales 
of length and velocity are arbitrary. }
\label{modes} 
\end{figure}

In order to highlight the essentials of non-normal amplification, we now 
take  a velocity field consisting of two modes, namely, a {\it spanwise streak} 
\beq 
{\bf u}_s = \left(\matrix{\alpha\sin\alpha y \cos\beta z\cr 0\cr 0}\right)\,, 
\eeq 
and a {\it downstream vortex} 
\beq 
{\bf u}_\omega =  
\left(\matrix{0\cr \beta \cos\alpha y \sin\beta z \cr -\alpha \sin\alpha y 
\cos\beta z}\right)\,, 
\eeq 
with amplitudes $s(t)\/$ and $\omega(t)\/$, i.e.,  
\beq 
{\bf u} = s(t) {\bf u}_s + \omega(t) {\bf u}_\omega \,. 
\eeq 
The two modes are sketched in Fig.~\ref{modes}. In the linearised equation the 
pressure disappears as the velocity fields are  divergence-free. The term  
$({\bf U}_0\cdot\nabla){\bf u}$ drops out and  
$({\bf u}\cdot\nabla){\bf U}_0$ results in a coupling between vortex and  
streak: 
\beq 
\left(\matrix{\dot s \cr \dot \omega}\right) = 
\left(\matrix{-\nu(\alpha^2+\beta^2) & S \cr 
0 & -\nu(\alpha^2+\beta^2) } \right) 
\left(\matrix{s \cr \omega}\right)\,. 
\label{matrix} 
\eeq 
The matrix on the right hand side is not symmetric because of the coupling of  
both modes through the term 
$({\bf u} \cdot \nabla) {\bf U}_0=S u_3 {\bf e}_x$.  
The dynamics that follows from the non-normal system 
(\ref{matrix}) has exponentially decaying vortices that drive spanwise 
streaks: if $s_0$ and $\omega_0$ denote the initial amplitudes, then 
\bea 
s(t) &=& (s_0 + S \omega_0 t) e^{-\nu(\alpha^2+\beta^2) t} ; \cr 
\omega(t) &=& \omega_0 e^{-\nu(\alpha^2+\beta^2) t} \,. 
\eea 
Clearly, even if there is no streak initially (i.e., $s_0=0\/$), there 
will be  
one as time progresses as a consequence of the mixing induced by the 
downstream  
vortex. Eventually, however, both will decay. The maximal amplitude of the  
streak follows from the maximum of $t \exp(-\nu(\alpha^2+\beta^2)t)$, which  
occurs at a 
time 
\beq 
t_{max} = \frac{1}{\nu (\alpha^2+\beta^2)} \,. 
\label{tmax} 
\eeq 
Since the maximal amplitude of the downstream component of the streak 
${\bf u}_s$ is $\alpha$, the maximal modulation of the  
downstream velocity component follows to be
\beq 
u_{s,max} = S \frac{\alpha}{\nu (\alpha^2+\beta^2)} \omega_0 \, . 
\label{umax} 
\eeq 
Note that this is proportional to the shear (the larger the shear the 
stronger  the amplification) and to the inverse of the viscosity.
This model also allows us to draw conclusions about the sizes of the vortices:  
Assume that the thickness in the shear direction, i.e., the wave number  
$\beta$, is given. Then the maximal value of $u_{s,max}$ is obtained  
from differentiation of Eq. (\ref{umax}) for  
$\alpha = \beta$. If $\beta=\pi/d\/$, 
with $d\/$ the separation between the  plates, the wavelength in the 
spanwise direction is about twice this width,  
i.e., $\lambda_{span} \simeq 2\pi/\beta=2d$.   

\subsection{Vortex-Shear Interactions with Noise}  

Suppose now that we add noise to the above system. 
Since the two components are  
already divergence-free, 
we model the noise by adding the random perturbations  
$\xi_s(t)$ and $\xi_\omega(t)$ to each component as follows: 
\beq 
\left(\matrix{\dot s \cr \dot \omega}\right) = 
\left(\matrix{-\lambda & S \cr 0 & -\lambda} \right) 
\left(\matrix{s \cr \omega}\right) + 
\left(\matrix{\xi_s \cr \xi_\omega}\right) \,. \label{matrix_noise} 
\eeq 
For simplicity we assume that 
\bea 
\langle \xi_s(t) \xi_s(t') \rangle &=& D_s \delta(t-t') \nonumber \\ 
\langle \xi_\omega(t) \xi_\omega(t') \rangle &=& 
D_\omega \delta(t-t') \nonumber \\
\langle \xi_s(t) \xi_\omega(t') \rangle &=& 0\,. 
\label{noise1} 
\eea 
The formalism is easily expanded to more general forms.
For notational convenience and since the quantitative dependence 
on wave numbers and viscosity is of minor  interest in this section, 
we have defined $\lambda = \nu ( \alpha^2+\beta^2)$. 
If we start at $t = -\infty\/$, the formal solution is 
\bea 
\omega(t) &=& \int_{-\infty}^t dt'\, e^{-\lambda (t-t')} \xi_\omega(t') ;  
\nonumber \\ 
s(t) &=& \int_{-\infty}^t dt'\, e^{-\lambda (t-t')} \xi_s(t') + \nonumber\\ 
&\ & S \int_{-\infty}^t dt'\, \int_{-\infty}^{t'}dt''\, 
e^{-\lambda (t-t')} e^{-\lambda (t'-t'')} \xi_\omega(t'') . 
\label{solution1} 
\eea 
With this choice of initial time any effects of initial conditions 
drop out and  the effects of noise are highlighted.  

\subsection{Correlations}  

\begin{figure}
\centerline{\epsfig{file=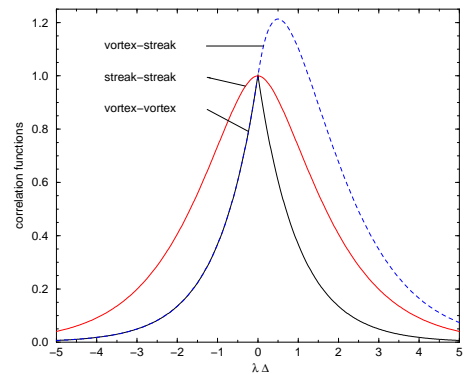,angle=-90,width=6cm}} 
\caption[]{ Correlation functions (see text) for vortex-streak 
coupling with noise of  
equal intensity in each component. The correlation functions are normalised by  
their values at zero, $\Delta\/$ is the time difference, and 
$\lambda\/$, 
which has dimensions of inverse time, is defined in the text.  
}\label{corrfcn} 
\end{figure}

The equations for the vorticity with their exponential driving 
and a  white-noise source represent an Ornstein-Uhlenbeck process 
and thus lead to  
exponentially decaying correlations \cite{Risken,Gardiner}. 
Formally this  follows from 
\bea 
&C&_{\omega,\omega}(t,\tau) = \langle \omega(t) 
\omega(\tau)\rangle = e^{-\lambda(t+\tau)}  \nonumber\\ 
&\times & \int_{-\infty}^t dt'  
\int_{-\infty}^\tau d\tau'\,  e^{\lambda(t'+\tau')} 
\langle \xi_\omega(t') \xi_\omega(\tau') \rangle \,, 
\eea 
which yields, in conjunction with Eq. (\ref{noise1})
 and the restriction of the  
final integral up to $\mbox{min}(t,\tau)\/$, 
\beq 
C_{\omega,\omega}(t,\tau) = \frac{D_\omega}{2\lambda}  e^{-\lambda |t-\tau|} . 
\label{cvv} 
\eeq  
The correlation function for the streaks has two components, 
one resulting from the noise in the streak components, and one 
from the non-normal amplification.  The cross terms involving 
the noise sources in both streak and vortex drop out  since 
they are uncorrelated. The streak contribution is again an  
Ornstein-Uhlenbeck process with an exponentially decaying 
correlation function  as in Eq. (\ref{cvv}).  The non-normal 
amplification gives rise to a contribution from the vortices, 
\bea 
&\ &C^{(vortex)}_{s,s}(t,\tau) =\int_{-\infty}^t dt' 
\int_{-\infty}^{t'} dt''   
\int_{-\infty}^\tau d\tau'  \int_{-\infty}^{\tau'} d\tau''  \nonumber\\ 
&\ &  
S^2 e^{-\lambda(t-t''+\tau-\tau'')}  
\langle \xi_\omega(t'') \xi_\omega(\tau'') \rangle . 
\eea 
Again via Eq. (\ref{noise1}) and the constraint on the 
domain of integration to  $\mbox{min}(t',\tau')\/$, 
the vortex contribution to the streak-streak  
correlation function becomes 
\beq 
C^{(vortex)}_{s,s}(t,\tau) =  
\frac{S^2 D_\omega}{4\lambda^3} 
\left(1+\lambda |t-\tau|\right)e^{-\lambda |t-\tau|}\,. 
\label{cssa} 
\eeq 
This shows clearly an algebraic contribution that results 
from the transient-growth characteristic of a non-normal 
system. For small  time-differences, the cusp at the 
origin for the Ornstein-Uhlenbeck process  becomes a 
rounded, quadratic maximum. The total correlation function  
\beq 
C_{s,s}(t,\tau) = \left( \frac{D_s}{2\lambda} +  
\frac{S^2 D_\omega}{4\lambda^3} 
\left(1+\lambda |\Delta|\right)\right) e^{-\lambda |\Delta|}\, , 
\label{css} 
\eeq 
with the time difference $\Delta \equiv t-\tau\/$, is symmetric in  
$\Delta\/$ and decays monotonically.  
Formally, this calculation is equivalent to that for a linearly damped 
process  exposed to both white noise (the driving of the streaks) 
and exponentially  correlated noise (the cross coupling from the vortices) 
\cite{Risken,Gardiner}.  What is specific to our case is that the 
coloured-noise component is generated  dynamically by the structure 
of the non-normal system.  Cross correlations between vortex 
and streak can be calculated similarly; and  they provide 
interesting insights into non-normal amplification. We define  
\beq 
C_{\omega,s}(t,\tau) = \langle \omega(t) s(\tau) \rangle\, . 
\eeq 
Then, for $t>\tau$, i.e., if the streak is probed {\it before} the vortex,  
cross correlations decay exponentially as  
\beq 
C_{\omega,s}(t,\tau) = \frac{S}{4\lambda^2}{D_\omega} e^{-\lambda(t-\tau)}  , 
\qquad 
\mbox{for}\ t>\tau \,.  
\label{csva} 
\eeq 
However, for $t<\tau$, i.e., if the streak is probed {\it after} the vortex,  
the driving of the streak by the vortex gives rise to the 
correlation function   
\bea 
C_{\omega,s}(t,\tau) &=& \frac{S}{4\lambda^2}{D_\omega}  
\left(1+2\lambda(\tau-t)\right) e^{-\lambda(\tau-t)} , \\ 
\label{csvb} 
&\ &\mbox{for}\ t<\tau \, , \nonumber 
\eea 
which increases first, for short times, and then decreases. It has a maximum  
that lies about $21\%$ above the value at zero, at $\lambda (\tau-t) =1/2$.  
The cross correlation thus provides a convenient and characteristic measure of  
the lift-up effect. A comparison between the three correlation functions  
discussed above is shown in Fig.~\ref{corrfcn}.  

\section{Noisy Shear Flows} 
\label{Fourier} 

In order to model noise in a shear flow, we have to allow for spatially  
fluctuating velocity fields. The dynamics we want to study, the interaction
between streaks and vortices, is mediated by the background shear flow.
While the detailed shape of either object and also the gradient of the
shear flow will depend on the specific boundary conditions, the conclusions
about the correlation function should be less sensitive. For instance,
the vortex-streak recycling mechanism is very similar in flows with
rigid boundary conditions and in those with free-slip boundary conditions
\cite{Waleffe1,Waleffe2}.
We, therefore, choose the theoretically most convenient form of
free slip boundary conditions so that we can use a 
Fourier representation for the velocity field, 
\beq 
{\bf u}({\bf x}, t) = \sum_{\bf k} {\bf u}({\bf k},t) e^{i{\bf k}\cdot {\bf x}} 
\eeq 
and also for the noise, 
\beq 
{\bf \xi}({\bf x}, t) = \sum_{\bf k} 
{\bf \xi}({\bf k},t) e^{i{\bf k}\cdot {\bf x}} \,.
\eeq 
The linearized Navier-Stokes equatoin (\ref{NS_lin}) becomes 
\begin{equation} 
\dot{{\bf u}}({\bf k},t) + S u_z({\bf k},t) \bf{e}_x =  
- \nu k^2 {\bf u}({\bf k},t) + {\bf \xi}({\bf k},t) \,. 
\label{Kelvineq} 
\end{equation} 
As in the previous section 
we assume that the wave vectors have no  downstream component.  
The stochastic force 
${\bf \xi}$ (or noise) satisfies 
\begin{eqnarray} 
\langle \xi_i({\bf k},t) \rangle &=& 0 \, , \\ 
\langle \xi_i({\bf k},t) \xi_j({\bf k}',t') \rangle 
&=&  A(k) P_{ij}({\bf k}) \delta({\bf k}+{\bf k}') 
\delta(t-t') \,, 
\end{eqnarray}  
where $A(k)\/$ is the amplitude of the variance (whose precise functional  
form does not matter for the moment but which will be 
discussed further at the end of this
section), 
$P_{ij}({\bf k}) = (\delta_{ij} - k_ik_j/k^2)$ 
is the transverse projector that enforces the incompressibility condition, and  
$i\/$ and $j\/$ 
denote Cartesian components. We assume, for simplicity, that  
the system is infinite; the effects of boundary conditions will be included  
presently by suitable constraints on $k_z\/$ as in the previous section.  
The temporal correlation functions we would like to calculate 
(cf. Sec.  \ref{vortex_streak}) are 
\beq 
C_{ij}(t,t') = \langle\langle u_i(t) u_j(t')\rangle\rangle\,, 
\label{cuu_def} 
\eeq 
where 
$\langle\langle \cdots \rangle\rangle$ stands for an integration over  
all points in the volume under consideration and an average over different  
realizations of the noise. Given our Fourier representation this becomes 
\beq 
C_{ij}(t,t') = \sum_{\bf k} \langle u_i({\bf k},t) u_j(-{\bf k},t') \rangle\,; 
\label{cuu_in_kspace} 
\eeq 
now only an average over different realizations of the noise remains  
for every wave number ${\bf k}\/$. This average can be calculated  
for each wave number individually and the sum can then be evaluated.  
Let us first write Eq. (\ref{Kelvineq}) in component form: 
\begin{eqnarray} 
\dot{u}_x({\bf k},t) + S u_z({\bf k},t) 
&=& - \nu k^2 u_x({\bf k},t)  + \xi_x({\bf k},t) ;  	\\
\dot{u}_y({\bf k},t) &=& - \nu k^2 u_y({\bf k},t) + \xi_y({\bf k},t) ; 	\\
 \dot{u}_z({\bf k},t) &=& - \nu k^2 u_z({\bf k},t) + \xi_z({\bf k},t) . 
\end{eqnarray} 
The projection onto divergence-free velocity fields and noise terms implies  
that the last two equations are coupled; they can thus be represented by a 
single field ${u}_\omega$ from which the $y\/$ and $z\/$ Cartesian  
components can be obtained as follows: 
\bea 
{u}_y({\bf k},t) &=& - u_\omega({\bf k},t) k_z/k ;\\ 
{u}_z({\bf k},t) &=& u_\omega ({\bf k},t) k_y/k\, . 
\eea 
Now only two equations remain for each ${\bf k}\/$: 
\bea 
\dot{u}_x({\bf k},t) + (S k_y/k) u_\omega({\bf k},t)  
&=& - \nu k^2 u_x({\bf k},t) + \xi_x({\bf k},t) ; 	
\label{eqnas}\\ 
\dot{u}_\omega({\bf k},t) &=& - \nu k^2 u_\omega({\bf k},t)  
+ \xi_\omega({\bf k},t)  \, ; 	
\label{eqnav} 
\end{eqnarray} 
and the noise terms satisfy 
\begin{equation} 
\langle \xi_a({\bf k},t) \xi_b({\bf k}',t') \rangle =  
A(k)\delta_{ab} \delta({\bf k}+{\bf k}') \delta(t-t') \,, 
\end{equation}  
where the indices $a\/$ and $b\/$ stand for, respectively, 
the noise components in downstream ($x$) and 
perpendicular direction (of vortex form, hence index $\omega$).  
Equations (\ref{eqnas}) and (\ref{eqnav}) are of the same form  
as Eq.~(\ref{matrix_noise}) so we can identify 
$\lambda\/$ with $\nu k^2$ and  
$S$ in Eq. (\ref{matrix_noise}) with 
$S\,k_y/k$ in Eq. (\ref{eqnas}). We can  therefore use the 
correlation functions (\ref{cvv}), (\ref{css}) and (\ref{csva}, \ref{csvb}) 
calculated previously 
along with the expansion of the 
velocity field (\ref{cuu_def}) and (\ref{cuu_in_kspace}) to obtain 
the correlation functions between spanwise and normal-velocity
component (subscripts $2$ and $3$, respectively),
\bea 
C_{2,2} =\sum_{\bf k}  \frac{k_z^2}{k^2} \langle u_\omega({\bf k}, t) 
u_\omega(-{\bf k}, t') \rangle\, , \label{c22}\\ 
C_{3,3} =\sum_{\bf k}  \frac{k_y^2}{k^2} \langle u_\omega({\bf k}, t) 
u_\omega(-{\bf k}, t')\rangle\, , \label{c33}\\ 
C_{2,3} =\sum_{\bf k}  \frac{-k_z k_y}{k^2} \langle u_\omega({\bf k}, t) 
u_\omega(-{\bf k}, t')\rangle\, , \label{c23} 
\eea 
in terms of  
\beq 
\langle u_\omega({\bf k}, t) u_\omega(-{\bf k}, t')\rangle = 
\frac{1}{2\nu k^2} A(k) e^{-\nu k^2 |t-t'|} \,. 
\eeq 
Thus these correlation functions decay exponentially; and unless there is an  
asymmetry in the weighting of wave numbers through the amplitude $A(k)\/$, 
the cross correlation $C_{2,3}$ vanishes on account of the  
antisymmetry under reflection of each wave number individually.  
The correlation function for the downstream component (subscript 1) becomes 
\beq 
C_{1,1}(t,t') = \sum_{\bf k} \frac{A(k)}{2\nu k^2} 
\left(1 + \frac{S^2 k_y^2}{2\nu^2 k^6}\left(1+\nu k^2 |\Delta|\right) \right) 
e^{-\nu k^2 |\Delta|}\,, 
\eeq 
where $\Delta=t-t'$ is the difference in times.  The cross correlations 
between vortex (as measured by the normal, $i=3$,  component of the velocity) 
and streak (measured by the downstream, $j=1$,  component) becomes 
\bea 
C_{3,1}(t,t') &=& \sum_{\bf k}  
\frac{S k_y^2}{\nu k^3}\frac{A(k)}{4\nu k^2} e^{-\nu k^2 (t-t')}, \\ 
&\ &\mbox{for}\ t>t' \, , 
\nonumber  
\eea 
and  
\bea C_{3,1}(t,t') &=& \sum_{\bf k}  
\frac{S k_y}{\nu k^3}\frac{A(k)}{4\nu k^2}  
(1+2\nu k^2(t'-t)) e^{-\nu k^2(t'-t)} , \\
 &\ &\mbox{for}\ t<t' \,.
\nonumber
\eea 
Interchanging $1$ and $3$ shows the
symmetry $C_{1,3}(t,t') = C_{3,1} (t',t)$.  

The spatial dependence or, equivalently, the wave-vector dependence of  the 
correlation functions is given by the summands in  
Eqs. (\ref{c22})-(\ref{c23}).  These correlation functions have high 
powers of the absolute value of $k$ in  the denominator, so all expressions 
are divergent for $k\rightarrow 0$, and  
thus dominated by the smallest wave numbers. 
In the absence of shear,  this divergence is much milder if we use the form of 
$A(k)\/$ appropriate for  thermal noise \cite{Onuki}. 
However, more singular forms have been used for $A(k)\/$ in an effort to 
obtain the Kolmogorov $k^{-5/3}\/$ energy spectrum at the  
level of a one-loop renormalization group \cite{RFNSE}. Of course the  
nonlinear term plays a crucial role in such calculations so it is not 
clear whether it is appropriate to use such singular forms for $A(k)\/$ 
in our study which does not include nonlinearities. 

So, unless the form of $A(k)\/$ favors a different selection, the
correlations will be dominated by the smallest wave numbers.
Given the bounding surfaces in the  $z$-direction, 
separated by a distance $d$,  
we have a normal wave number $k_z=n \pi/d$,  
with $n$ a non-vanishing integer, and a continuous spanwise wave number $k_y$.  
The smallest value of $n\/$ is $n=1\/$ and the associated optimal  
$k_y$ is $n\pi/d$. Rigid boundary conditions may change these 
scales somewhat.

\section{Concluding Remarks} 
\label{conclusions}  

We have discussed correlations in noisy shear flow, both at the  
level of a simple model with two degrees of freedom and in a model  
with spatially varying velocity fields. The main result is that
the non-normal coupling between downstream vortices and 
streaks is reflected in the temporal cross correlation function
between the downstream and normal velocity components. 

While there have been 
many studies of spatial correlations (Townsend \cite{Townsend} summarizes
a large body of literature), few have addressed temporal correlations
and hardly any the cross correlations we described here.
For instance, Farrell et al. \cite{Farrell2} study only the variance 
maintained  by forcing a noisy shear flow. 
Onuki \cite{Onuki} concentrates on the  case in which $A(k)\/$ has 
the form appropriate for thermal noise and then  calculates the 
renormalisation of the viscosity because of the nonlinear term.  
Bassam et al \cite{Bassam} concentrates on energy amplification 
but not on the  correlation functions we consider.
Blackwelder and Kovasznay \cite{Blackwelder} measured the
cross correlations but their diagrams do not provide sufficient
detail at short times.

It is, therefore, only in recent numerical work on 
Lagrangian models \cite{RDT} that the Lagrangian version of this
cross correlation has been considered \cite{Pope2}. Analysis of 
Lagrangian data shows indeed the asymmetry in the cross
correlation due to the non-normal effects. Eulerian cross correlations
have be studied in a numerical simulation of 
turbulent shear flow \cite{Schumacher} and they also show
the cross correlations and their asymmetry in time
\cite{Princeton}. 

These numerical simulations also indicate that the 
relation of our
linear model with independent, additive and white in time
random perturbations to natural fluctuations in a turbulent flow
is not as far fetched as it may seem. 
Certain time scales and quantitative details will be affected
by the nonlinearities neglected in our calculation,
but if there is an overall shear the main mechanism
of non-normal growth will always be present,
and that is at the heart of the cross correlations we study here.

Many of the features of the cross correlations we 
discuss should be accessbile experimentally. The exponential  
correlation in the vortices should be reflected in the correlations of  
transversal, i.e., spanwise and normal, velocity components. The streaks 
are a  property of the downstream velocity component and the long 
correlations  should be visible there. Finally, the streak-vortex 
correlations should be noticeable in longitudinal-transversal 
correlations. We are not aware of experimental measurements 
of such correlations. The most direct test would involve fluctuations
in laminar shear flows. Two experimental set ups come to mind: 
\begin{enumerate} 
\item  One could exploit the inevitable free-stream turbulence in  
a wind tunnel as a noise source. A shear gradient can then be 
imposed by a flat plate, above which a linear shear gradient 
forms. Crossed hotwire anemometers can be used to characterise 
the statistics of the upstream noise and the velocity field above 
the plate. The free decay of the noise between the point of 
characterisation and the point of measurement should have a 
negligible effect. Reynolds numbers should  be low enough so 
that no turbulence transitions in the shear layer are  induced.  
\item There are several experiments or proposals for driving and 
influencing flows in electrically conducting fluids by electrical 
and magnetic fields \cite{Tabeling,Gollub,Brosa}. Random 
perturbations of well defined strength could be induced  
by random fluctuations in the currents. Ideally, flow geometries 
that are closed, such as a Taylor-Couette apparatus, or spatially 
localized, as in a plane Couette flow \cite{Alfredsson,Dauchot} 
should be used. Such a setup should allow for a good control 
of the noise and should avoid complications arising from the  
advection of perturbations.  
\end{enumerate}  

Finally, we would like mention that there is growing interest  
in turbulent transport in magnetohydrodynamics in the presence of  
a background shear \cite{MHD}. The techniques developed here can  
be applied fruitfully to this case as well, as we will show elsewhere  
\cite{BrunoMHD}.  

\section*{Acknowledgements} This research was 
started while we were visiting the 
Institute of  Theoretical Physics, University 
of California at Santa Barbara, USA, where  
it was supported in part by the US National Science 
Foundation under Grant  
No. PHY94-07194. For further support BE would like to thank the  
Deutsche Forschungsgemeinschaft and RP the Indo-French Centre for the  
Promotion of Advanced Research (Project No. 2404-2). We would also like  
to thank Sriram Ramaswamy for discussions.     

  
\narrowtext  

\end{multicols} 

\begin{references}   
\bibitem[\ast]{byjnc} 
Also at Jawaharlal Nehru Centre for Advanced Scientific Research, 
Bangalore, India.    

\bibitem{Grossmann} S. Grossmann, {Rev. Mod. Phys.} {\bf 72}, 603 (2000).  

\bibitem{Landahl} M. Landahl, {J. Fluid Mech.} {\bf 98}, 243 (1980).  

\bibitem{Boberg}  L. Boberg and U. Brosa, 
{ Z. Naturforsch. A} {\bf 43}, 697 (1988).  

\bibitem{Trefethen} N.L. Trefethen, A. Trefethen, 
S.C. Reddy and T.A. Driscoll, { Science} {\bf 261}, 578 (1992).  

\bibitem{Farrell1} B.F. Farrell and P.J. Ioannou, 
{ Phys. Fluids A} {\bf 5}, 2298 (1993).  

\bibitem{Ecology} M.G. Neubert and H. Caswell, 
{Ecology} {\bf 78}, 653 (1997). 

\bibitem{SH} P.J. Schmid and D.S. Henningson, 
{\em Stability and 
transition in shear flows}, (Springer, New York, 2001).
 
\bibitem{Farrell2} B.F. Farrell and P.J. Ioannou, 
{Phys. Rev. Lett.} {\bf 72}, 1188 (1994).  

\bibitem{Farrell3} P.J. Ioannou, 
{J. Atmos. Sci.} {\bf 51}, 998 (1995).  

\bibitem{Farrell4} B.F. Farrell and P.J. Ioannou, 
{Phys. Fluids A} {\bf 5}, 1390 (1993).  

\bibitem{Onuki} A. Onuki, {Physics Letters}, {\bf 70A}, 31 (1979). 

\bibitem{Hamilton} J.M. Hamilton, J. Kim, and F. Waleffe,
{J. Fluid Mech.} {\bf 287}, 317 (1995).

\bibitem{Waleffe1} F. Waleffe,
{Phys. Fluids} {\bf 7}, 3060 (1995).

\bibitem{Waleffe2} F. Waleffe,
Phys. Fluids {\bf 9}, 883 (1997).

\bibitem{RDT} S. Pope, 
{\em Turbulent Flows} (Cambridge University Press, Cambridge, 2000). 

\bibitem{Nazarenko}
S. Nazarenko, N.K.-R. Kevlahan and B. Dubrulle, 
Physica D {\bf 139}, 158 (2000).

\bibitem{Kelvin} Lord Kelvin, 
{ Philos. Mag.} {\bf 23}, 459 (1887).  

\bibitem{Craik} A.D.D. Craik and W.O. Criminale, 
{Proc. R. Soc. (London), A} {\bf 406}, 13 (1986).  

\bibitem{Marzinzik} B. Eckhardt, J. Kronj\"ager and K. Marzinzik, 
Nonnormal amplification in viscoelastic fluids, to be submitted.

\bibitem{Risken} H. Risken, 
{\em The Fokker-Planck Equation} (Springer, Berlin, 1996).  

\bibitem{Gardiner} C.W. Gardiner, 
{\em Handbook of Stochastic Methods} (Springer, Berlin, 1996). 

\bibitem{RFNSE} See, e.g., A. Sain, Manu, and R. Pandit,  
{Phys. Rev. Lett.}, {\bf 81}, 4377 (1998) and references therein. 

\bibitem{Townsend}
A.A. Townsend, {\em The structure of turbulent shear flows},
2nd ed., (1976), (Cambridge University Press, Cambridge).

 \bibitem{Bassam}  B.~Bamieh and M.~Dahleh, 
{\it Energy Amplification in  Channel Flows with 
Stochastic Excitation}, preprint (1999) 
[http: //online.itp.ucsb.edu/online/hydrot00/bamieh/]. 

\bibitem{Blackwelder}
R.F. Blackwelder, and L.S. Kovasznay,
Phys. Fluids {\bf 15}, 1545 (1972).

\bibitem{Pope2}
S.B. Pope,
Phys. Fluids {\bf 14}, 1070 (2002).

\bibitem{Schumacher}
J. Schumacher, and B. Eckhardt, 
Europhys. Lett. {\bf 52}, 627 (2000).

J. Schumacher, J. Fluid Mech. {\bf 441}, 109 (2001).

\bibitem{Princeton}
B. Eckhardt, J. Schumacher and A. Jachens, 
{\em Asymmetric time correlations in turbulent shear flows}.
Proceedings IUTAM meeting on turbulent
shear flows, Princeton 2002.
 
\bibitem{Tabeling} O. Cardoso, D. Marteau and P. Tabeling, 
Phys. Rev. E{\bf 49}, 454 (1994).

\bibitem{Gollub} B.S. Williams, D. Marteau and J. P. Gollub, 
Phys. Fluids {\bf 9}, 2061 (1997).

\bibitem{Brosa} U. Brosa, { Z. Naturforsch.} {\bf 46a}, 473 (1991).  

\bibitem{Alfredsson} N. Tillmark and P.H. Alfredsson, 
{ J. Fluid Mech.} {\bf 235}, 89 (1992).  

\bibitem{Dauchot} F.~Daviaud{,} J. Hegseth{,}~P. Berg\'e. 
{ Phys. Rev. Lett.} {\bf 69}, 2511 (1992).  

\bibitem{MHD} E-j Kim and B. Dubrulle, 
{ Phys. Plasmas},  {\bf 8}, 813 (2001).  

\bibitem{BrunoMHD} B. Eckhardt and R. Pandit, to be published. 

\end{references}
\end{document}